# Predicting Age from White Matter Diffusivity with Residual Learning

Chenyu Gao[a,*], Michael E. Kim[b], Ho Hin Lee[b], Qi Yang[b], Nazirah Mohd Khairi[a], Praitayini Kanakaraj[b], Nancy R. Newlin[b], Derek B. Archer[c,d], Angela L. Jefferson[c,d,e], Warren D. Taylor[f], Brian D. Boyd[g], Lori L. Beason-Held[h], Susan M. Resnick[h], The BIOCARD Study Team[1], Yuankai Huo[b], Katherine D. Van Schaik[i], Kurt G. Schilling[i], Daniel Moyer[b], Ivana Išgum[j], Bennett A. Landman[a,b,d,f,i]

[1]Data used in preparation of this article were derived from BIOCARD study data, supported by grant U19 –AG033655 from the National Institute on Aging. The BIOCARD study team did not participate in the analysis or writing of this report, however, they contributed to the design and implementation of the study. A listing of BIOCARD investigators may be accessed at: https://www.biocard-se.org/public/Core%20Groups.html

[a]Dept. of Electrical and Computer Engineering, Vanderbilt University, Nashville, USA; [b]Dept. of Computer Science, Vanderbilt University, Nashville, USA; [c]Vanderbilt Memory and Alzheimer's Center, Vanderbilt University Medical Center, Nashville, USA; [d]Dept. of Neurology, Vanderbilt University Medical Center, Nashville, USA; [e]Dept. of Medicine, Vanderbilt University Medical Center, Nashville, USA; [f]Dept. of Psychiatry and Behavioral Sciences, Vanderbilt University Medical Center, Nashville, USA; [g]Vanderbilt Center for Cognitive Medicine, Vanderbilt University Medical Center, Nashville, USA; [h]Laboratory of Behavioral Neuroscience, National Institute on Aging, National Institutes of Health, Baltimore, USA; [i]Dept. of Radiology and Radiological Sciences, Vanderbilt University Medical Center, Nashville, USA; [j]Dept. of Biomedical Engineering and Physics, Dept. of Radiology and Nuclear Medicine, Amsterdam University Medical Center, University of Amsterdam, Amsterdam, Netherlands

## ABSTRACT

Imaging findings inconsistent with those expected at specific chronological age ranges may serve as early indicators of neurological disorders and increased mortality risk. Estimation of chronological age, and deviations from expected results, from structural magnetic resonance imaging (MRI) data has become an important proxy task for developing biomarkers that are sensitive to such deviations. Complementary to structural analysis, diffusion tensor imaging (DTI) has proven effective in identifying age-related microstructural changes within the brain white matter, thereby presenting itself as a promising additional modality for brain age prediction. Although early studies have sought to harness DTI's advantages for age estimation, there is no evidence that the success of this prediction is owed to the unique microstructural and diffusivity features that DTI provides, rather than the macrostructural features that are also available in DTI data. Therefore, we seek to develop white-matter-specific age estimation to capture deviations from normal white matter aging. Specifically, we deliberately disregard the macrostructural information when predicting age from DTI scalar images, using two distinct methods. The first method relies on extracting only microstructural features from regions of interest (ROIs). The second applies 3D residual neural networks (ResNets) to learn features directly from the images, which are non-linearly registered and warped to a template to minimize macrostructural variations. When tested on unseen data, the first method yields mean absolute error (MAE) of 6.11 ± 0.19 years for cognitively normal participants and MAE of 6.62 ± 0.30 years for cognitively impaired participants, while the second method achieves MAE of 4.69 ± 0.23 years for cognitively normal participants and MAE of 4.96 ± 0.28 years for cognitively impaired participants. We find that the ResNet model captures subtler, non-macrostructural features for brain age prediction.

*chenyu.gao@vanderbilt.edu; phone 1 667 910-5300; https://my.vanderbilt.edu/masi/people/

## 1. INTRODUCTION

Each person's brain ages in its own unique trajectory, emphasizing the need for a precise biomarker that gauges the "true" biological age of a brain, relative to chronological age. Studies have shown that large deviations between chronological and biological age can indicate conversion from mild cognitive impairment (MCI) to Alzheimer's disease (AD)[1], brain atrophy after traumatic injury[2], schizophrenia[3], increased mortality risk[4], major depressive disorder[5,6], and other brain disorders[7]. Existence of a general biomarker for biological brain age holds wide-reaching implications, as it could serve as a tool in guiding clinical interventions for brain diseases and disorders.

Diffusion tensor imaging (DTI) facilitates a non-invasive exploration of the degree of anisotropy and structural orientation based on water movements within the architecture of the tissues.[8] It provides a depth of information about brain microstructure beyond the capabilities of normal structural MRI modalities such as T1-weighted MRI.[9] Studies have shown that brain microstructure experiences age-related alterations throughout the lifespan,[10] offering insights into normal brain aging,[11] as well as abnormal brain aging like Alzheimer's disease.[12] Fractional anisotropy (FA) and mean diffusivity (MD) are two commonly used scalar maps derived from DTI data. In cognitively normal participants, we observe changes in brain microstructure and diffusivity as age increases (Figure 1). We would like to know whether we can use this information for age prediction. Specifically, we study brain age prediction methods that do not depend on macrostructural differences. We minimize these differences through non-rigid and non-linear registrations between participants and a target atlas space, which involves warping of the brain anatomy. All inferences are made using only FA and MD within the normalized space. In this context, we are not seeking to develop the most accurate brain age prediction possible. Instead, we aim to identify and characterize the aspects of aging that impact white matter microstructure. The overall goal is to develop a metric that is complementary to more traditional structural brain age prediction approaches[13].

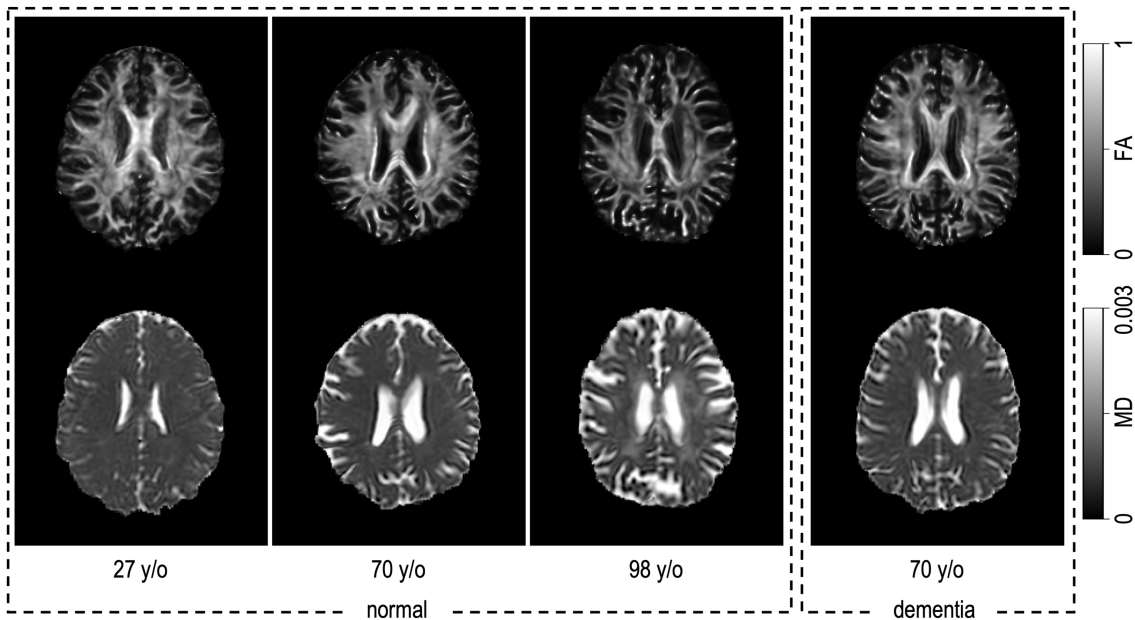

Figure 1. The premise of this effort is that the brain undergoes microstructural and diffusivity changes throughout the normal aging process. At left above, we can appreciate that there are microstructural changes—as shown in fractional anisotropy (FA)—mainly characterized by a decrease in FA (top row). Additionally, there are diffusivity changes, with increased diffusivity in the white matter, notably in the central white matter, as shown in mean diffusivity (MD) (bottom row). We would like to know if prediction of the chronological age from microstructure and diffusivity could provide a useful biomarker to detect abnormal aging as a difference between the age one might predict from a participant with dementia (shown right), versus their true chronological age.

## 2. METHODS

We propose two distinct methods for predicting brain age from DTI scalar images. (Figure 2) The first method involves whole-brain segmentation and the extraction of features from each region of interest (ROI). These extracted features are then input into a multi-layer perceptron (MLP), which generates the predicted age. We adopt this method as our baseline, providing us with a benchmark for minimum achievable performance using a straightforward approach. The second method leverages a 3D ResNet[14] to learn features directly from the images. The high-dimensional features are then fed into an MLP to yield the predicted age. To determine the best-performing models, we use 5-fold cross-validation, with consistent fold-splitting across all models. Finally, we assess the performance of models of both methods on the same, previously unseen testing sets, using images of cognitively normal participants and cognitively impaired participants.

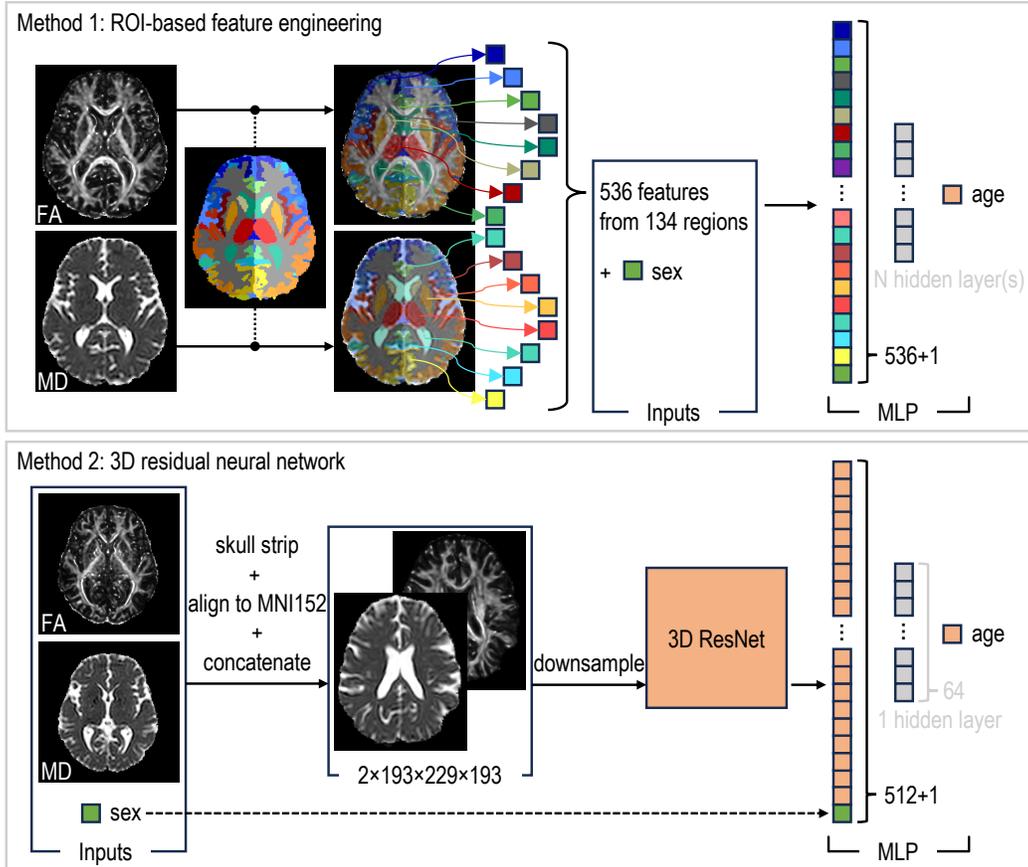

Figure 2. The ROI-based feature engineering method uses mean and standard deviation values of FA and MD within each ROI (segmented by SLANT[16,17]), alongside the sex of the participant, to feed into an MLP. The 3D ResNet method extracts features from preprocessed images. These features, once concatenated with the participant's sex, are then processed by an MLP, with or without a hidden layer, to generate a prediction of the participant's age.

### 2.1 Data

The FA and MD images are generated from DTI data preprocessed by the PreQual[15] pipeline. We use brain segmentation labels created by SLANT[16,17], aligning them with the brain presented in the FA and MD images, to extract the mean and standard deviation values of FA and MD within each ROI (Figure 2), and to generate binary brain masks for removing non-brain voxels in the FA and MD images. We aim to reduce the macrostructural information that the 3D ResNet models can learn from the FA and MD images. To achieve this, we align and warp the brain in these images to match with the one in the MNI152 template[18], by performing a series of both linear and non-linear registrations implemented by ANTs[19].

All resulting images are manually inspected, and those with unsuccessful preprocessing are excluded. After inspection, we have a dataset composed of 1327 participants (Table 1). The collected data is split at the participant-level into training (which contains only cognitively normal participants) and testing sets (which contains both cognitively normal and impaired participants). The training set is then divided into five consecutive folds. Each fold, in turn, is used once as the validation set while the remaining four folds constitute the training set. The data splits are exported to .csv files, and we ensure no participant overlap occurs between the training, validation, and testing sets.

Table 1. We use data acquired from 1327 participants for the training, validation, and testing of the models. There are mean chronological age shifts across datasets, which makes the prediction more challenging on the unseen testing data.

| Site | Training + Validation | | Testing (normal participants) | | Testing (impaired participants) | |
|---|---|---|---|---|---|---|
| | # Participants | Mean Age | # Participants | Mean Age | # Participants | Mean Age |
| BIOCARD[23] | 104 | 68.7 ± 8.4 | 35 | 69.6 ± 7.2 | 84 | 73.9 ± 8.4 |
| BLSA[24] | 895 | 65.1 ± 14.7 | 117 | 72.5 ± 13.0 | 72 | 82.9 ± 7.4 |
| ICBM[25] | 19 | 28.0 ± 5.7 | 1 | 40 | 0 | N/A |
| Total | 1018 | 64.7 ± 15.0 | 153 | 71.6 ± 12.2 | 156 | 78.1 ± 9.1 |

## 2.2 ROI-based feature engineering method

We extract the mean and standard deviation of FA and MD values from 134 ROIs, along with the sex of the participant, resulting in a total of 537 features. These features are fed as input to an MLP (Figure 2). We experiment with different configurations of hidden layers, each varying in the number of neurons.

## 2.3 3D residual neural network

We build and train the 3D ResNets with PyTorch[20] and MONAI[21], using a Quadro RTX 5000 with 16 GB of RAM. The FA and MD images, each sized at 193×229×193 at 1 mm isotropic voxels, are combined (with each image being treated as a separate channel) and resampled to 128×128×128 at 1.51×1.79×1.51 mm anisotropic voxels before being input into the ResNet. The ResNet subsequently generates 512 features from each FA and MD image pair. These 512 features are combined with sex information and provided to an MLP, which generates the predicted brain age (Figure 2). We experiment with multiple ResNet architectures of varying model complexities, as well as MLPs with and without the hidden layer between the input and output layers.

## 2.4 Model evaluation

Through a 5-fold cross-validation process, we identify the optimal model of each method and evaluate its generalizability and applicability using two unseen testing sets. One set comprises cognitively normal participants, while the other includes cognitively impaired participants (not MCI), participants with mild cognitive impairment (MCI), and participants with dementia.

## 3. RESULTS

For the ROI-based feature engineering method, the MLP configuration with layers arranged as (input→128→64→output) outperforms all other configurations from this method, achieving MAE of 6.31 ± 0.48 years on the validation set. Despite this, every 3D ResNet configuration, even the one with the poorest performance on the validation set, the ResNet10 concatenated with an MLP that has one hidden layer, is significantly better (t-statistic=3.83, p-value=0.019) than the best of the ROI-based feature engineering method across validation folds (Figure 3). For the 3D ResNets, increased complexity does not necessarily lead to improved performance (compare ResNet34 with ResNet18 in Figure 3). Also, adding an extra hidden layer to the MLP does not always improve performance (Figure 3). ResNet18 concatenated with an MLP that has one hidden layer yielded the best performance (MAE=4.85 ± 0.16 years) across the validation folds.

On the unseen testing sets, the best model from the ROI-based feature engineering method achieves MAE of 6.11 ± 0.19 years on cognitively normal participants and MAE of 6.62 ± 0.30 years on cognitively impaired participants, while the best model among the 3D ResNets achieves MAE of 4.69 ± 0.23 years on cognitively normal participants and MAE of 4.96 ± 0.28 years on cognitively impaired participants. Upon completion of the training, validation, and testing of these two chosen models, we assessed the performance of all models on the testing sets. In this post-hoc comparison, our chosen ResNet model was the second-best regarding MAE on the testing sets (Table 2). For clarification, the age predictions presented in the subsequent figures are made by the MLP (input→128→64→output) (the lowest purple triangle in Figure 3) trained on the 5[th] fold or ResNet18 trained on the 5[th] fold (the lowest blue circle with black outline in Figure 3).

The kernel density of the difference between predicted and chronological age (Figure 4) shows that: (i) Compared to the ResNets, the ROI-based feature engineering method has larger differences between predicted age and chronological age; (ii) For cognitively normal participants, while the majority of age predictions are centered around the chronological age (zero difference), the spread of these predictions is quite broad, and there are a few outliers for both methods; (iii) On cognitively impaired participants, the differences between predicted and chronological age tend to deviate from zero, and there is a noticeable trend of the density distribution becoming increasingly narrower and more skewed as the severity of impairment advances.

On cognitively normal participants sampled from the testing set, we visually observed that brain changes correlate more strongly with increases in predicted age than with increases in chronological age (Figure 5). Macrostructural changes, such as ventricle enlargement, are not the major influences in the prediction. Rather, subtler, non-macrostructural features seem to guide our brain age prediction model.

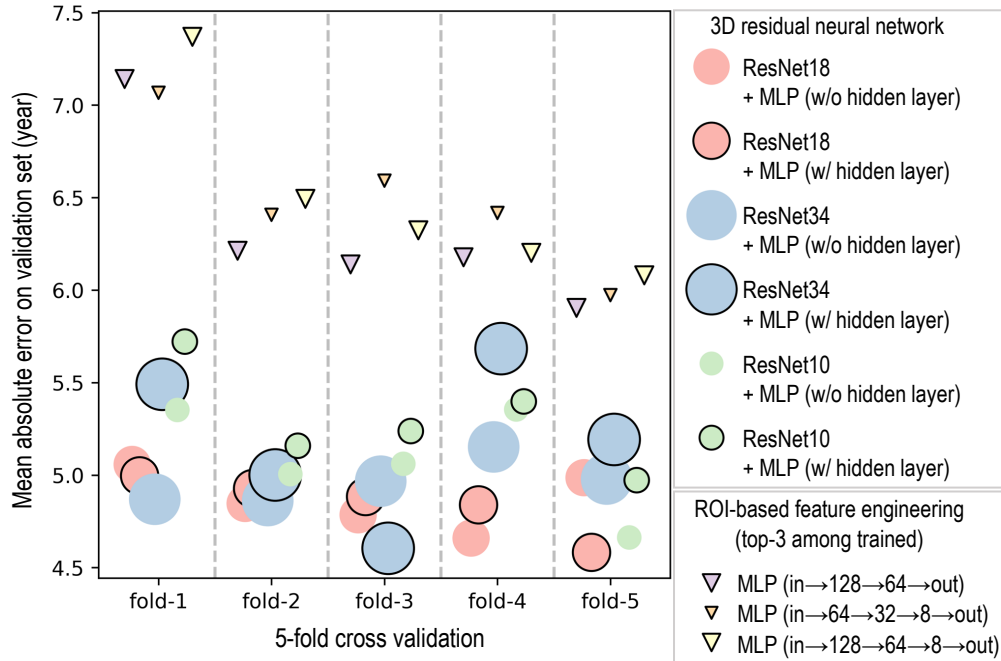

Figure 3. The best model from the ROI-based feature engineering method, MLP with layers arranged as (input→128→64→output), is significantly worse than the worst model from the 3D ResNet method, ResNet10 concatenated with an MLP that has one hidden layer (t-statistic=3.83, p-value=0.019). ResNet18 (with hidden layer in the MLP) is significantly better (t-statistic=2.905, p-value=0.044) than ResNet10 (with hidden layer in the MLP), and better (t-statistic=2.347, p-value=0.079) than ResNet34 (with hidden layer in the MLP). Having a hidden layer in the MLP does not necessarily improve the performance. For instance, ResNet10 (with hidden layer in the MLP) is significantly worse (t-statistic=3.629, p-value=0.022) than ResNet10 (without hidden layer in the MLP). The size of the markers represents the relative complexity of the models (as measured by the number of trainable parameters) in comparison to other models from the same method. (All t-statistic and p-value presented here are calculated from paired t-test)

## 4. DISCUSSION

We aimed to predict white matter age by focusing solely on microstructural and diffusivity features, refraining from using macrostructural features. To achieve this, we used a combination of linear and non-linear registrations to align the input images to a standard template and warp the anatomy to match the one in the template image. This helps to normalize the macrostructures within the images, thereby minimizing the macrostructural information the model can use for age prediction. Even with the macrostructural information minimized (if not completely removed), our best model shows a performance comparable to that reported in existing literature. For instance, Chen et. al reported that their cascade neural network model– which takes tract features extracted from 76 fiber tract bundles as input– achieves MAE of 4.78 years on the unseen data after refined optimization and transfer learning[22], while our best model achieves MAE of 4.69 years on the unseen data. But note that our model was trained on a larger training set (N=800 for each fold, compared to N=500 for theirs). And we did not test our model performance on data coming from different sites like they did. We also note that it is not our goal to develop the most accurate brain age prediction possible in this project. Instead, the overall goal is to develop a metric that is complementary to more traditional structural brain age prediction approaches. As we integrate more data into our training set and begin to include macrostructural features, along with other types of data, we anticipate the potential to fully leverage the predictive capabilities of our model.

Table 2. The 3D ResNets achieve lower MAE than the ROI-based feature engineering method. ResNet18 concatenated with an MLP (with one hidden layer of 64 neurons) achieves the best performance on the validation set and the 2nd-best on the unseen testing sets. The lowest MAE on the validation set for each method is highlighted in bold.

| Method | Model | Validation (normal) MAE (year) | Testing (normal) MAE (year) | Testing (impaired) MAE (year) |
|---|---|---|---|---|
| ROI-based feature engineering | MLP (in→128→64→out) | **6.31 ± 0.48** | 6.11 ± 0.19 | 6.62 ± 0.30 |
| | MLP (in→64→32→8→out) | 6.49 ± 0.39 | 6.33 ± 0.23 | 6.90 ± 0.36 |
| | MLP (in→128→64→8→out) | 6.49 ± 0.51 | 6.43 ± 0.27 | 6.99 ± 0.40 |
| 3D residual neural network | ResNet10 + MLP (w/o hidden layer) | 5.09 ± 0.29 | 5.15 ± 0.23 | 5.41 ± 0.23 |
| | ResNet10 + MLP (w/ hidden layer) | 5.30 ± 0.28 | 5.28 ± 0.27 | 5.70 ± 0.28 |
| | ResNet18 + MLP (w/o hidden layer) | 4.87 ± 0.16 | 4.48 ± 0.13 | 4.86 ± 0.15 |
| | ResNet18 + MLP (w/ hidden layer) | **4.85 ± 0.16** | 4.69 ± 0.23 | 4.96 ± 0.28 |
| | ResNet34 + MLP (w/o hidden layer) | 4.97 ± 0.12 | 4.81 ± 0.17 | 4.96 ± 0.38 |
| | ResNet34 + MLP (w/ hidden layer) | 5.19 ± 0.42 | 4.99 ± 0.14 | 5.35 ± 0.29 |

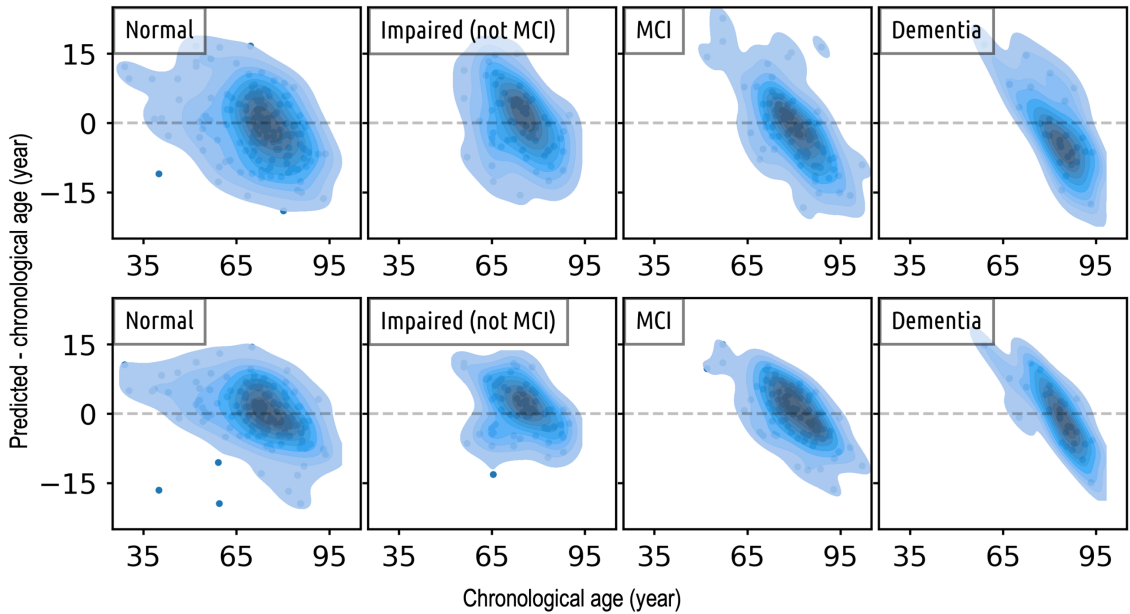

Figure 4. Cross-sectional comparison of age predictions made by the top-performing model from both the ROI-based feature engineering approach (top row) and the 3D ResNet method (bottom row) shows that: i) In comparison to the predictions made by the ROI-based feature engineering method, the ages predicted by the ResNet model generally align more closely with the chronological ages; ii) As we transition from normal, to impaired, to MCI, and finally, to dementia, the density distribution becomes increasingly narrower and more diagonal. This pattern suggests a trend of diminishing model performance as the severity of the disease increases.

# 5. CONCLUSION

Identification of deviations from the typically-expected changes that occur with progression of chronological age is crucial for the early detection and diagnosis of neurological pathology. We developed models using two distinct methods for predicting white matter specific age, relying exclusively on the microstructural and diffusivity information present in DTI scalar images. Notably, we minimized the use of macrostructural information in these models. Our experimental findings affirm that DTI data can serve as valuable input for predicting brain age.

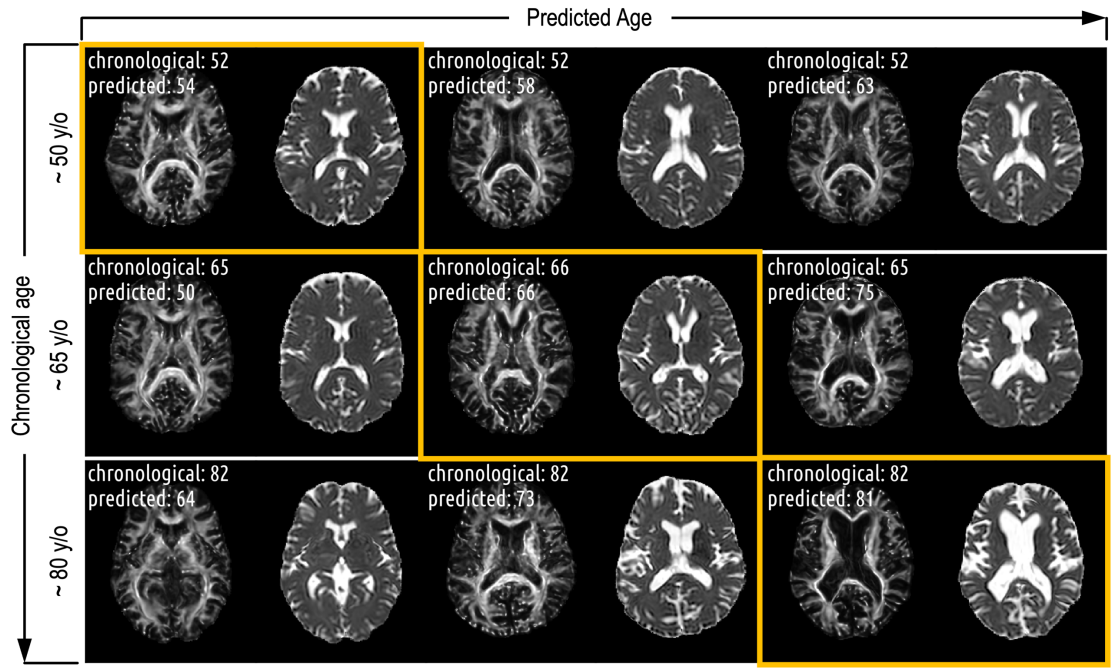

Figure 5. There are more pronounced changes of the brain along the predicted age axis compared to the chronological age axis. While macrostructural changes (those remained after the registrations) like ventricle enlargement apparently influence age prediction in older participants, they are not the primary factors driving the age prediction, particularly for participants with predicted ages under 70. In these cases, subtler features, distinct from apparent macrostructural changes, are guiding the prediction.

## ACKNOWLEDGMENT

This work was supported by the National Institutes of Health under award numbers R01EB017230, 1K01EB032898, K01-AG073584 and T32GM007347, and in part by the National Center for Research Resources, Grant UL1 RR024975-01, and UL1-TR000445, S10-OD023680 (Vanderbilt's High-Performance Computer Cluster for Biomedical Research) and U24-AG074855. This work was conducted in part using the resources of the Advanced Computing Center for Research and Education at Vanderbilt University, Nashville, TN. The Vanderbilt Institute for Clinical and Translational Research (VICTR) is funded by the National Center for Advancing Translational Sciences (NCATS) Clinical Translational Science Award (CTSA) Program, Award Number 5UL1TR002243-03. The content is solely the responsibility of the authors and does not necessarily represent the official views of the NIH.

The BLSA is funded by the Intramural Research Program of the National Institute on Aging, NIH.

The BIOCARD study is supported by a grant from the National Institute on Aging (NIA): U19-AG03365. The BIOCARD Study consists of 7 Cores and 2 projects with the following members: (1) The Administrative Core (Marilyn Albert, Corinne Pettigrew, Barbara Rodzon); (2) the Clinical Core (Marilyn Albert, Anja Soldan, Rebecca Gottesman, Corinne Pettigrew, Leonie Farrington, Maura Grega, Gay Rudow, Rostislav Brichko, Scott Rudow, Jules Giles, Ned Sacktor); (3) the Imaging Core (Michael Miller, Susumu Mori, Anthony Kolasny, Hanzhang Lu, Kenichi Oishi, Tilak Ratnanather,